\documentclass[twoside,a4paper,11pt]{sca}
\usepackage{graphicx}
\usepackage{hyperref}
\usepackage{movie15}
\usepackage{natbib}  
\newcommand{\Teff} {T$_{\rm eff}$}
\newcommand{\grav} {log\,{\em g}}

\begin{document}
\pagenumbering{arabic}
\pagestyle{myheadings}
\thispagestyle{empty}
\vspace*{1.7cm}
\begin{flushleft}
{\bf {\LARGE
%
The IACOB project: synergies for the Gaia era 
%
}\\
\vspace*{1cm}
%
S. Sim\'on-D\'iaz$^{1,2}$, M. Garcia$^{1,2}$, A. Herrero$^{1,2}$,
J. Ma\'iz Apell\'aniz$^{3}$, \& I. Negueruela$^{4}$ 
%
}\\
\vspace*{0.5cm}
%
$^{1}$
Instituto de Astrof\'isica de Canarias, E-38200 La Laguna, Tenerife, Spain.\\
$^{2}$
Departamento de Astrof\'isica, Univ. de La Laguna, E-38205 La Laguna, Tenerife, Spain.\\
$^{3}$
Instituto de Astrof\'isica de Andaluc\'ia, CSIC, Apdo. 3004, E-18080 Granada, Spain.\\
$^{4}$
Dpto. de F\'isica, Ingenier\'ia de Sistemas y teor\'ia de la Se\~nal. Escuela 
Polit\'ecnica Superior. Univ. Alicante, Apdo. 99, E-03080 Alicante, Spain.\\
%
\end{flushleft}
%
\markboth{
The IACOB project: synergies for the Gaia era 
}{ 
%
S. Sim\'on-D\'iaz et al.%
}
\thispagestyle{empty}
\vspace*{0.4cm}
\begin{minipage}[l]{0.09\textwidth}
\ 
\end{minipage}
\begin{minipage}[r]{0.9\textwidth}
\vspace{1cm}
\section*{Abstract}{\small
The {\em IACOB spectroscopic survey of Galactic OB stars} is an ambitious observational
project aimed at compiling a large, homogeneous, high-resolution database of optical
spectra of massive stars observable from the Northern hemisphere. The 
quantitative spectroscopic analysis of this database, complemented by the 
invaluable information provided by Gaia (mainly regarding photometry and distances), 
will represent a major step forward in our knowledge of the fundamental physical 
characteristics of Galactic massive stars. In addition, results 
from this analysis will be of interest for other scientific questions to be 
investigated using Gaia observations. In this contribution we outline the present status 
of the {\em IACOB spectroscopic database} and indicate briefly some of the synergy links between the IACOB 
and Gaia scientific projects.
%
\normalsize}
\end{minipage}
%
%
%
\section{Introduction \label{intro}}

The Gaia mission will gather a complete astrometric, photometric and spectroscopic 
dataset of over one billion stars. Its instrument performance \citep[see e.g.,][]{Prusti11} 
is designed to provide information 
about proper motions, radial velocities, photometric variability, distances, 
interstellar reddening, rotational velocities, atmospheric parameters, and element 
abundances of the observed stars. The Gaia observations will hence definitely 
challenge many aspects of our present view of the stellar component in the Milky Way.

\begin{figure}[t!]
\centering
\includegraphics[height=15.cm , angle=90]{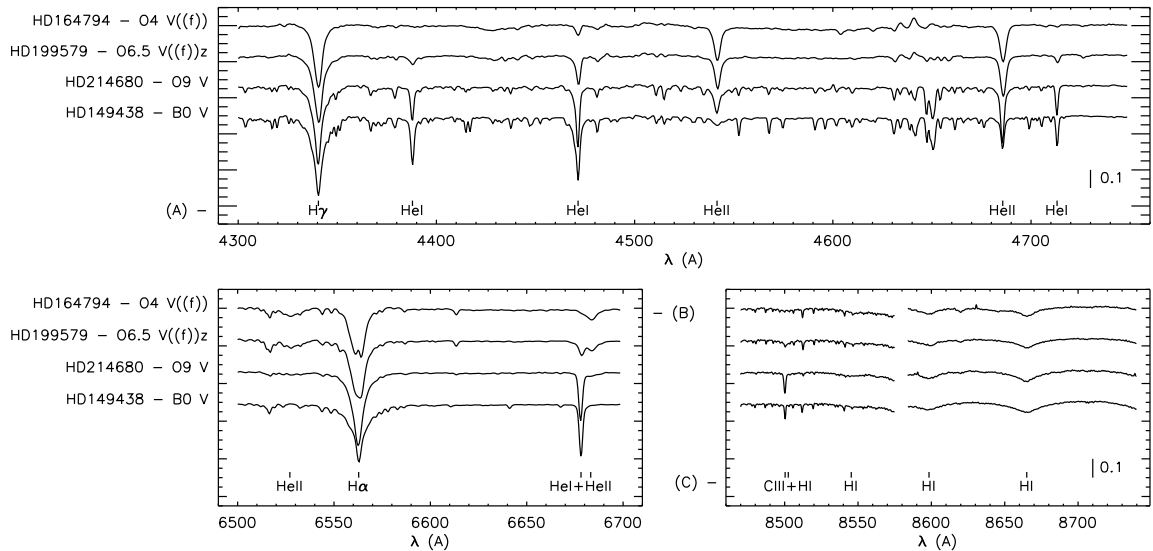}
\caption{{\small {\em IACOB-sweg} spectra of four representative O stars in the optical 
(panels A and B) and Gaia (panel C) spectral ranges. Contrary to the optical range,
the Gaia range is almost empty of stellar features in the case of O and early B-type
stars. Note that the small narrow lines in the Gaia range (appart from the C\,{\sc ii} line) 
are telluric.
\label{fig1}}}
\end{figure}

While Gaia photometric and spectroscopic observations will allow a proper characterization 
of the stellar properties of late-B and cooler stars (including abundances), a similar 
study in the case of O and early-B type stars will be somewhat limited. On the one
hand, the Balmer jump (planned to be observed by the Blue Photometer) is less accurate
as indicator of the stellar temperature in this type of stars than in the case
of cooler stars. On the other hand, as illustrated in Figure \ref{fig1}, the spectral 
range observed by the Radial Velocity Spectrometer is almost empty of spectral features 
which could be used to determine rotational velocities, and stellar parameters/abundances 
(contrarily to the optical range). Although Gaia observations 
will allow to distinguish between early/mid-O, late-O, and early-B type stars, a proper
(accurate) determination of the stellar parameters will require to combine these observations
with optical spectra (see more details in Section \ref{analysis}).

Complementing the Gaia database with optical spectroscopic observations of O and early-B stars
is hence of crucial importance for any Gaia scientific project involving stellar parameters and/or 
abundances of massive stars. Some examples in the context of this conference include the full 
characterization of the IMF of stellar clusters (also accounting for the high-mass tail), the study 
of the impact of massive stars on the formation and dynamical evolution of young 
stellar clusters and associations, or the investigation of self-pollution by SNe Type II 
products in stellar associations with signatures of triggered star formation.

Although originally motivated by scientific drivers not directly related to 
Gaia\footnote{The GREAT (Gaia Research for European Astronomy 
Training) networking programme also plans to observe massive stars in selected clusters as part of more 
general long term spectroscopic surveys preparing and following Gaia \citep[see][]{Blo11}.}, 
large (on going) optical spectroscopic surveys of O and B-type stars such as GOSSS (P.I. Ma\'iz Apell\'aniz), 
OWN (P.I. Barb\'a), NoMaDS (P.I. Pellerin), IACOB (P.I. Sim\'on-D\'iaz), and IACOB-sweg (P.I. Negueruela)
will become very valuable in the Gaia era. All these surveys, initially independent, are now coordinated
and the corresponding spectroscopic databases are planned to be made public in the future for their
maximum exploitation in different scientific contexts.

This contribution concentrates on the {\em IACOB spectroscopic database} and its scientific exploitation, 
specially remarking some of the synergy links between the IACOB and Gaia scientific projects. 
Notes on the other surveys can be found elsewhere \citep[e.g.][]{Bar10, Sot11, Mai11}.

\section{The IACOB project}

The {\em IACOB project} (P.I. Sim\'on-D\'iaz) aims at progressing in our knowledge of 
Galactic massive stars using a large, homogeneous, high-quality spectroscopic dataset 
and modern tools for the quantitative spectroscopic analysis of O and B-type stars. 
The project is divided into six main working packages. The first one is devoted to the
compilation of the {\em IACOB spectroscopic database of Northern Galactic OB stars},
a very important piece of the project. The other five refer to the scientific exploitation
of this unique database of high-resolution spectra, covering several aspects of interest 
in the massive star research:
\begin{table}[ht] 
\center
\begin{tabular}{l l l  l  l } 
{\bf WP-1:} &  The {\em IACOB spectroscopic database} & & {\bf WP-4:} &  Abundances in OB-type stars \\
{\bf WP-2:} &  Line-broadening in OB stars &  & {\bf WP-5:} &  Binary/multiple systems \\
{\bf WP-3:} &  Quantitative spectroscopic analyses & & {\bf WP-6:} &  massive stars and the ISM \\
\end{tabular} 
\label{tab1} 
\end{table}
\\ \newline
The number of studies planned to be developed in the framework of the IACOB project (i.e. using spectra from the
IACOB database) is large. At present, we are concentrating our efforts on the investigation of the so-called 
{\em macroturbulent} broadening affecting the line-profiles of massive stars and its possible connection to 
stellar pulsations \citep[e.g.][]{Sim10b}, on the revision of rotational velocities of Galactic OB stars, 
and on the accurate determination of the stellar and wind parameters of the single O and B stars in the sample. 
Some IACOB spectra have also been used to investigate the degree of chemical (in)homogeneity of B-type 
stars in the Ori\,OB1 association \citep{Sim10a, Nie11}, or to provide a firm comfirmation of the presence of 
a third massive star component in the $\sigma$\,Ori\,AB system \citep{Sim11b}. \\
\newline
Gaia observations will have a definite importance in some of the outcomes of the IACOB project (mainly regarding
WP3). In addition, the analysis of spectra from the IACOB database (in combination with Gaia observations) will 
provide unvaluable information for other scientific projects studying our Galaxy and/or some of its components. Some 
examples of the synergies between IACOB and Gaia will be provided in Sect. \ref{analysis}, but first 
we briefly summarize the present status and some future prospects of the {\em IACOB database}.

\subsection{The IACOB spectroscopic database: updates and future prospects\label{database}}

The first spectra for the {\em IACOB spectroscopic database} were obtained with the FIES instrument attached to 
the Nordic Optical Telescope (NOT) in El Roque de los Muchachos Observatory (La Palma, Spain) in 
November 2008. Since then, and up to August 2011, we have compiled $\sim$\,1000 spectra of 
over 200 bright Galactic OB stars with $\delta\,\ge\,$-25\,deg. The present version of the database 
(v2.0) is mainly built on the IACOB v1.0 \citep{Sim11a}
plus the following updates: (1) the number of observed stars and compiled multi-epoch spectra has 
increased during 14 new observing nights from 105 and 720 to 204 and 984, respectively; (2) the 
database is now complete up to V\,=\,8 mag in the case of O-type stars; (3) the number of observed 
B0\,--\,B2 stars has also increased (now also including giants, and not only dwarfs and supergiants).

New observations are already schedulled with FIES for September 2011. We plan (1) to observe fainter
O stars (up to V = 9 mag) with a somewhat lower resolution (R\,=\,23\,000), (2) to increase the number of 
B-type stars in the database, (3) to obtain at least 3 epochs for the whole sample, and (4) 
to follow up some of the detected binary/multiple systems. 

Since June 2011 (and planned to be completed in November 2011) the {\em IACOB database} also
counts on a new supplement ({\em IACOB-sweg}, P.I. I. Negueruela). These new observations, obtained with the HERMES 
spectrograph attached to the Mercator telescope in La Palma, extend further to the red (up to 9000 \AA) than the 
FIES spectra. They were planned to investigate the spectroscopic behavior of MK standars with spectral types 
earlier than B9 in the Gaia range for a optimal future exploitation of the GAIA spectra of this type of stars.

\subsection{Quantitative spectroscopic analyses of massive stars in the Milky Way \label{analysis}}

As discussed in Sect. \ref{intro}, and illustrated in Figure \ref{fig1} the Gaia spectroscopic
observations will be empty of spectral features which could be used to extract information 
about the fundamental physical parameters of OB-type stars. The situation is different
when we consider the optical spectral range. The quantitative spectroscopic analysis of this 
spectral range makes it possible to determine the rotational velocities, effective temperatures,
gravities and abundances of these stars \citep[see e.g. the recent review by][and references 
therein]{Mar11}. When the optical analyses are complemented with the information provided by 
other spectroscopic diagnostics (mainly the UV) we can also constrain the wind properties of the 
analyzed stars. 

Nowadays we count with very powerful tools, such as advanced stellar atmosphere codes and 
stellar evolution models, which have allowed us to make a remarkable progress in our 
knowledge of the physical and wind properties of massive stars \citep[see review by][]{Pul08}. 
The outcome of the quantitative spectroscopic analysis of medium and high resolution optical spectra
of Galactic OB-type stars by means of modern stellar atmosphere codes is now accepted to 
have reached a high degree of reliability \citep[viz.][]{Her02, Rep04, Lef07, Mar08, Sim10a, Nie10}. 
However, it is important not to forget that our interpretation of these results may be clearly biased 
in some cases, mainly due to observational limitations. 

The first point refers to the stellar luminosities, radii and masses. To obtain these fundamental
parameters we need to combine the derived spectroscopic parameters with information about the distance 
and extinction to the star. At present, the distance to most massive stars in our Galaxy are poorly
constrained and hence luminosities, radii and masses resulting from our analyses are somewhat 
uncertain in many cases. Table \ref{10Lac} illustrates how critical is this point in the case of the 
O9\,V star 10\,Lac (d$\sim$580 pc).  Fortunately, this situation will change very 
soon thanks to Gaia. 

The second point refers to binarity/multiplicity. Most or all massive stars are suspected to be 
born as part of multiple systems \citep[see e.g., ][and references therein]{Zin07, Mas09}. Many 
of these binary/multiple systems remain still unresolved; therefore, it would not be so strange
to find studies of massive stars in which its multiple nature has not been detected (since
these are based on single-epoch spectroscopic and photometric observations) and consequently 
the derived stellar parameters are erroneous.

Finally, even in isolated objects, other secondary properties such as mass-loss rate and 
rotation also have an important impact on stellar evolution of massive stellar objects 
\citep{Mey00}; while it would be desirable to consider a sample of isolated stars large 
enough to investigate in a systematic manner the effect of all these parameters, the number of objects 
analyzed up to now in a consistent way is still limited.

Quantitative spectroscopic analyses of multi-epoch, large sample observations of Galactic OB stars 
(combining spectroscopy, photometry, and accurate distance determinations) are hence one of the
next important steps to improve our knowledge of massive stars and clarify some of the gaps 
that still remain in this important field of Astrophysics. The {\em IACOB project} is following 
this approach in the case of Galactic stars.
At present, we are performing the analysis of the stars included in the {\em IACOB spectroscopic database} 
following an automatic best-fit strategy based on a extensive grid of synthetic HHe spectra computed 
with the FASTWIND stellar atmopshere code \citep[see details in][]{Sim11c}. As a results of this analysis, 
the project will provide a complete database of stellar parameters 
of Galactic OB stars homogeneously determined. As commented above, the use of
accurate distances and multi-band photometry will be of crucial importance to complete the stellar
parameters determined spectroscopically with luminosities, radii, and masses. In this sense, Gaia
observations will come right on time.
%
\begin{table}[t!]
\centering
\caption{\label{10Lac}Results from the analysis of 10\,Lac (O9\,V) assuming the present accuracy
in distance, photometry and stellar parameters. The uncertainties in $R$, $L$, and $M$ due
to error propagation of the uncertainties in stellar parameters (\Teff\ and \grav) and distance
are indicated separately. The later one (in brackets) clearly dominates. We need more
accurate distances.}
\begin{tabular}{r l |l l}
\noalign{\smallskip}
\hline
\noalign{\smallskip}
$d$\,=589$\pm$79 pc (13\%) & \cite{Meg09} & \Teff\,=\,35000$\pm$600 K & \\ 
$m_{\rm v}$\,=4.879$\pm$0.008 & \cite{Mai04} & \grav\,=\,3.87$\pm$0.10 dex& \\
($B-V$)\,=\,-0.201$\pm$0.010 & \cite{Mai04} & \\
($B-V$)$_0$\,=\,-0.27 & \cite{Mar06} &  $R$/$R_{\odot}$\,=\,7.52$\pm$0.09 & [$\pm$1]\\
$A_{\rm v}$\,=\,0.214$\pm$0.03 &  & log\,($L$/$L_{\odot}$)\,=\,4.88$\pm$0.02 & [$\pm$0.12]\\
$M_{\rm v}$\,=\,-4.15$\pm$0.30 &  & $M$/$M_{\odot}$\,=\,16$\pm$3 & [$\pm$5] \\
\noalign{\smallskip}
\hline
\noalign{\smallskip}
\end{tabular}
\end{table}


The final aim of the IACOB spectra and the associated database of stellar parameters (including 
also rotational velocities and abundances) is to increase the amount of observational 
constraints which may help to advance in some of the still outstanding problems in the field of 
massive stars. In addition, they can be used to study e.g. the kinematics of some of the 
Galactic OB clusters and associations, or the impact that massive stars have on the 
formation of other stellar objects (via their strong winds and ionizing fluxes).
Finally, we plan to provide the spectral energy distributions (SEDs) predicted by the 
FASTWIND code for all the analyzed stars so they can be then used in the 
study of the interstellar extinction along different line of sights (when combined 
with the observed SEDs provided by Gaia).

\newpage
%
\small  
%
\section*{Acknowledgments}   
%
SSD kindly acknowledge the staff at the Nordic Optical Telescope for their
professional competence and always useful help during the observing nights.
Financial support by the Spanish Ministerio de Ciencia e
Innovaci\'on under projects 
AYA2008-06166-C03-01 and	
AYA2010-21697-C05-05.
AYA2010-21697-C05-04,
and by the Gobierno de Canarias under project 
PID2010119.
This work has also been partially funded by the Spanish MICINN under the 
Consolider-Ingenio 2010 Program grant CSD2006-00070: First Science with the GTC
({\tt http://www.iac.es/consolider-ingenio-gtc}).%

%


\end{document}